\documentclass[aps,twocolumn,amssymb,showpacs]{revtex4}

\usepackage{epsfig}
\usepackage[centertags]{amsmath}
\usepackage{amsfonts}
\usepackage{amssymb}

\begin{document}

\title{Study of transfer reaction channel in $^{12}$C + $^{27}$Al system}

\author{Aparajita Dey\footnote{Email: aparajita@vecc.gov.in}, M.~Biswas, C.~Bhattacharya, T.~K.~Rana, S.~Kundu, K.~Banerjee, S.~Bhattacharya, T.~K.~Ghosh, G.~Mukherjee, J.~K.~Meena, D.~Gupta\footnote{Present address: Bose Institute, Department of Physics and
Centre for Astroparticle Physics and Space Science, Block EN, Sector V, Salt Lake City, Kolkata - 700 091, India.}, P.~Mali\footnote{Department of Physics, University of North Bengal, Siliguri - 734013, India.}, S.~Mukhopadhyay, D.~Pandit, H.~Pai, S.~R.~Banerjee}

\affiliation{Variable Energy Cyclotron Centre, 1/AF, Bidhan Nagar, Kolkata - 700064, India}

\author{Suresh Kumar, A.~Srivastava, A.~Chatterjee, K.~Ramachandran, K.~Mahata, S.~Pandit, S.~Santra}

\affiliation{Nuclear Physics Division, Bhabha Atomic Research Centre, Trombay, Mumbai - 400085, India}

\begin{abstract}
The 1p transfer channel in the $^{27}$Al($^{12}$C, $^{11}$B)$^{28}$Si reaction has been studied at E$_{lab}$ = 73, 81 and 85 MeV. The finite range distorted wave Born approximation calculations have been performed using phenomenological optical model potential to analyze the angular distributions for 3 transitions populating the 0.0, 1.78 and 4.62 MeV states of $^{28}$Si and 2 transitions populating the 2.12 and 4.44 MeV states of $^{11}$B via the $^{27}$Al($^{12}$C, $^{11}$B)$^{28}$Si reaction. The spectroscopic strengths as well as spectroscopic factors have been extracted for all the five states. The extracted strength values are compared with shell model calculations.
\end{abstract}


\maketitle

\section{Introduction}

The study of single nucleon transfer reaction gives the valuable information on nuclear structure. In recent years, there has been a lot of interest in studying transfer reactions, and heavy ion induced reactions provide a wide opportunity for studying various transfer channels \cite{brog77,scott74}. One specific feature of heavy-ion induced reactions is the possibility to observe the excitations of both fragments in the exit channel. For example, in the $^{27}$Al($^{12}$C, $^{11}$B)$^{28}$Si reaction, the excited states of $^{28}$Si and $^{11}$B may be seen in the $^{11}$B energy spectrum. The finite range distorted-wave Born approximation (DWBA) has been extensively used to describe the low-energy single-nucleon ($e.g.$, one proton, 1p or one neutron, 1n) transfer reactions \cite{wil75,thor75,bec75,hud75,far06,gui07}.

Previous studies of heavy-ion induced transfer reactions have enormous uncertainties arising from the ambiguities in the optical model potential. The strong absorption in the heavy-ion elastic scattering has restricted its sensitivity to the extreme surface region of the nucleus, and a variety of potentials is known to provide good fits to the experimental data, provided that they have similar values in this critical (extreme surface) region. However, for higher bombarding energies, the potential is probed over a wider domain inside the strong absorption radius \cite{bar86}. There are enough evidence that the phenomenological optical model potentials can be quite precisely determined, at least for relatively light projectiles and targets \cite{bran88,bran2,fric88}.

In the present work, we have studied the 1p transfer reaction channel in the $^{27}$Al($^{12}$C, $^{11}$B)$^{28}$Si reaction at 73, 81 and 85 MeV bombarding energies. Earlier, this reaction was studied at 46.5 MeV \cite{po75} and $E/A$ = 50 MeV \cite{win89} bombarding eneries, however the information about spectroscopic factors was not given properly. The ground state spectroscopic strength was determined using two types of optical model potential and the values are widely varied (2.4 and 7.7) \cite{win89}. Kalifa {\it et al.} reported a value of 2.9 \cite{kal73} for the ground state spectroscopic strength.  In addition to these, the strength values for different states of $^{28}$Si were extracted many times using the light ion induced reaction and they are found to be different \cite{bar68,das99}. On the other hand, the spectroscopic strengths for different states of $^{11}$B were reported in literature \cite{dev79,war04} and there are disagreement between the results.

It is well known that at incident energies in the region of 10 MeV/nucleon, the angular momentum carried by projectiles, such as, $^{12}$C or $^{16}$O on light nuclei usually exceeds the critical angular momentum which the compound system can support. Consequently, the direct aspects of the reaction mechanism should be enhanced in this energy region. In the present study, the bombarding energies are in the region 6 -- 7 MeV/nucleon and the Coulomb barrier of the system is 21.23 MeV. The excitation energies of the composite system are in the range 67 -- 75 MeV. Therefore, we can observe different excited states of target-like and projectile-like fragments in the spectra.

The article is arranged as follows. The experimental details are given in the next section. The experimental results are presented in Sec.~III and discussed. Finally, the summary and conclusion is given in Sec.~IV.

\section{Experimental Details}

The experiments were performed using $^{12}$C and $^{11}$B ion beams from the BARC-TIFR 14UD pelletron accelerator at Mumbai. During the first experiment, the $^{12}$C ion beams, having energies 81 and 85 MeV, were bombarded on a self-supporting $^{27}$Al target of thickness $\sim$ 550 $\mu$g/cm$^2$. Both elastic and transfer channels were detected using Si-Si ($\sim$ 10 $\mu$m Si $\Delta$E and $\sim$ 350 $\mu$m Si E) and Gas-Si ($\sim$ 80 Torr P10 $\Delta$E and 450 $\mu$m Si E) telescopes. The two types of telescopes were mounted on two arms of the scattering chamber which could move independently. Typical solid angle subtended by Si-Si telescope was $\sim$ 6.4 msr and Gas-Si telescope was $\sim$ 4.5 msr. The well separated ridges corresponding to different fragments are clearly seen in $\Delta$E--E scatter plot. The angular distributions of elastic scattering and 1p transfer channel have been taken in a wide angular range. The telescopes were calibrated using elastically scattered $^{12}$C ion from Al target and Th-$\alpha$ source.

\begin{figure} [h] \begin{center}
\epsfysize=11.0cm
\epsffile[63 57 474 763]{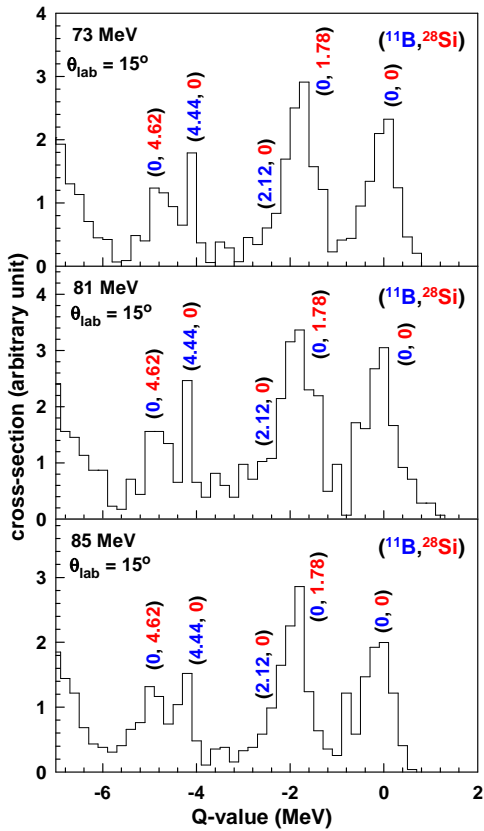}
\end{center}
\caption{(colour online) The partial energy spectrum of $^{11}$B fragment emitted in $^{27}$Al($^{12}$C, $^{11}$B)$^{28}$Si reaction at 73, 81 and 85 MeV bombarding energies at an angle $\theta_{lab}$ = 15$^o$. The different states are labeled according to the excitation energies (in MeV) in $^{11}$B and $^{28}$Si.}
\label{fig1}
\end{figure}

In the second experiment, the $^{12}$C ion beam of energy 73 MeV was bombarded on the same Al target and the angular distributions of elastic scattering and 1p transfer channel have been measured in a wide angular range. In addition, the $^{11}$B ion beam of energy 64.4 MeV was bombarded on a Si target of thickness $\sim$ 420 $\mu$g/cm$^2$. The elastic scattering data have been taken in a wide angular range. The same set of telescopes were used to detect the emitted particles; however, typical solid angle subtended by the Gas-Si telescope was $\sim$ 14.4 msr and Si-Si telescope was $\sim$ 12.6 msr in this experiment.  The telescopes were calibrated using elastically scattered $^{12}$C ion from Al target and $^{11}$B ion from Si target and Th-$\alpha$ source. The systematic errors in the data, arising from the uncertainties in the measurements of the solid angle, target
thickness, and the calibration of current digitizer have been
estimated to be $\approx$ 15\% in both the experiments. 

\section{Results and Discussions}

The energy spectrum of the outgoing $^{11}$B fragment obtained at an angle $\theta_{lab}$ = 15$^o$ is shown in Fig.~\ref{fig1} for 73, 81 and 85 MeV bombarding energies. In the
energy spectrum, the ground state (Q-value = $-$ 4.372 MeV) and different excited states of $^{11}$B itself and $^{28}$Si have been identified and indicated by their energies in MeV. 

The angular distribution have been analysed for the ground state (0$^+$; $E_X$ = 0.0 MeV), first excited state of $^{28}$Si (2$^+$; $E_X$ = 1.78 MeV), second excited state of $^{28}$Si (4$^+$; $E_X$ = 4.62 MeV), first excited state of $^{11}$B (1/2$^-$; $E_X$ = 2.12 MeV) and second excited state of $^{11}$B (5/2$^-$; $E_X$ = 4.44 MeV).

\begin{figure} [h] \begin{center}
\epsfysize=10.0cm
\epsffile[43 198 470 760]{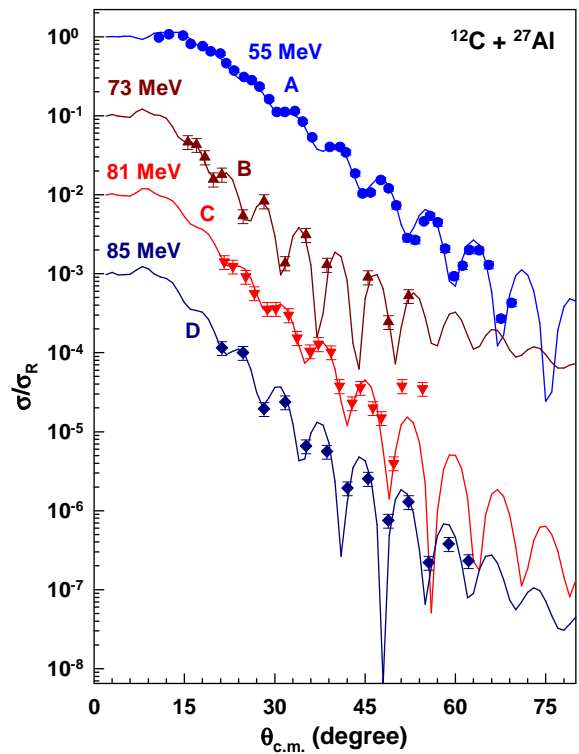}
\end{center}
\caption{(colour online) Elastic scattering angular distributions for $^{12}$C + $^{27}$Al reaction at 55 (data taken from Ref.~\cite{park77}) (circle), 73 (triangle), 81 (inverted triangle) and 85 (square) MeV. The solid lines represent the optical model fitting using parameters given in Table~\ref{tbl1}.}
\label{op1}
\end{figure}

\subsection{Optical model potential}
The optical model parameters have been derived from the elastic scattering data for both entrance and exit channels. We performed the optical model analysis using the parametric Woods-Saxon (WS) forms for both the real and imaginary potentials. The phenomenological optical model potential which describe the elastic angular distribution at each energy has the following form
\begin{eqnarray*}
U(r) = U_{OM}(r) + V_C(r) ~~~~~~~~~~~~~~~~~~~~~~~~~~~~~~~~~~ \\
U_{OM}(r) = -V(r; V_o, R_o, a_o) ~~~~~~~~~~~~~~~~~~~~~~~~~~~~~ \\
- i[ W_F(r; W_v, R_v, a_v)  + W_D(r; W_s, R_s, a_s) ]
\end{eqnarray*}

\noindent
where $V(r)$ denotes the volume type WS real potential, $W_F(r)$ is a volume type WS imaginary potential to simulate the fusion after penetration of the barrier, $W_D(r)$ is a derivative type WS imaginary potential to account for the absorption due to reactions at the surface and $V_C(r)$ is the Coulomb potential.

\begin{table*} 
\caption{Best fit parameters with phenomenological potential. }
\begin{eqnarray*}
U(r) = \frac{-V_o}{1 + \exp(r - R_o)/a_o} + i \left[ \frac{-W_v}{1 + \exp(r - R_v)/a_v} + \frac{-W_s}{1 + \exp(r - R_s)/a_s} \right] + V_C(r)  \\
R_i = r_i(A_P^{1/3} + A_T^{1/3})~~~ (i = o, v, s), ~~r_C = 1.31 \hbox{fm}~~~~~~~~~~~~~~~~~~~~
\end{eqnarray*}
\begin{tabular}{ccccccccccccc} \\ \hline\hline
Energy&Set&$V_o$&$r_o$&$a_o$&$W_v$&$r_v$&$a_v$&$W_s$&$r_s$&$a_s$&$\chi^2/N$&$\sigma_r$ \\
(MeV)&&(MeV)&(fm)&(fm)&(MeV)&(fm)&(fm)&(MeV)&(fm)&(fm)&&(mb) \\ \hline\hline
\multicolumn{13}{c}{$\bf ^{12}C + ^{27}Al$} \\
55&A&27.32&2.212&0.57&36.32&1.921&0.67&1.08&1.120&0.67&26&1563 \\
73&B&27.32&2.207&0.57&33.32&1.881&0.67&1.13&1.120&0.67&32&1627 \\
81&C&19.32&1.961&0.57&22.32&1.731&0.67&1.18&1.120&0.67&33&1284 \\
85&D&19.32&1.955&0.57&21.52&1.680&0.67&1.18&1.120&0.67&41&1238 \\ \hline
\multicolumn{13}{c}{$\bf ^{11}B + ^{28}Si$} \\
49.5&E&27.16&1.914&0.66&34.80&1.823&0.66&5.65&1.609&0.78&35&1521 \\ 
64.4&F&27.16&1.914&0.66&20.80&1.723&0.66&5.64&1.609&0.78&45&1500 \\ \hline\hline
\end{tabular}
\label{tbl1}
\end{table*}

The fitting procedure can be summarised as follows: The search code ECIS94 \cite{ecis} was used to perform the optical model calculations to obtain the parameters of the best fit potential. The initial parameters were taken from Refs.~\cite{park77} and \cite{park79} for entrance channel and exit channel, respectively. These potentials have volume real and imaginary potential terms, however, derivative type imaginary potential term was not included in the calculations. We introduce this term in our present calculations.

\subsubsection{Entrance channel}
In the present work, the elastic scattering angular distributions for the $^{12}$C + $^{27}$Al reaction have been measured at energies 73, 81 and 85 MeV. Previously, this system was studied at various energies \cite{park77,roy79,glo80} and 55 MeV was the highest incident energy found in the literature. At this energy the potential parameters were $V(r)$(32.31, 1.23, 0.57) and $W_F(r)$(27.71, 1.12, 0.67) and $r_C$ = 1.31 fm \cite{park77}. We use these parameters as the starting parameters. In the first search, the volume imaginary potential $W_F$ and real radius $R_o$ were kept fixed and searches were performed over the remaining five free parameters, viz., $V_o$, $a_o$, $W_s$, $R_s$ and $a_s$. Subsequently, changing the $W_v$, $R_v$, $a_v$ and $R_o$ in steps, same search was repeated again to obtain the best fit parameters with minimum $\chi^2$ value. The final set of best fit parameters for 55 MeV (data taken from Ref.~\cite{park77}) corresponding to minimum $\chi^2/N$ value are given in Table~\ref{tbl1} (Set A). For the other incident energy ($e.g.$, 73 MeV) the same search procedure was followed with the best fit parameters of 55 MeV as the starting parameter set. The best fit parameters with minimum $\chi^2/N$ value for 73 MeV are given in Table~\ref{tbl1} (Set B). Subsequently, Set B was taken as the starting parameters to obtain the best fit parameters for 81 MeV (Set C). Again for 85 MeV Set C was taken as the starting parameters. The best fit paramters for 85 MeV are given in Table~\ref{tbl1} (Set D). It has been found that the resultant geometry parameters along with the strengths of the potential components are energy dependent. For instance, the radius values of both the real and imaginary components decreases with increasing energy. The best fit parameters, the minimum $\chi^2/N$ values and the corresponding reaction cross-sections $\sigma_r$ have been given in Table~\ref{tbl1}. The fits are shown in Fig.~\ref{op1} by solid lines.

\begin{figure} [h] \begin{center}
\epsfysize=7.0cm
\epsffile[47 344 470 760]{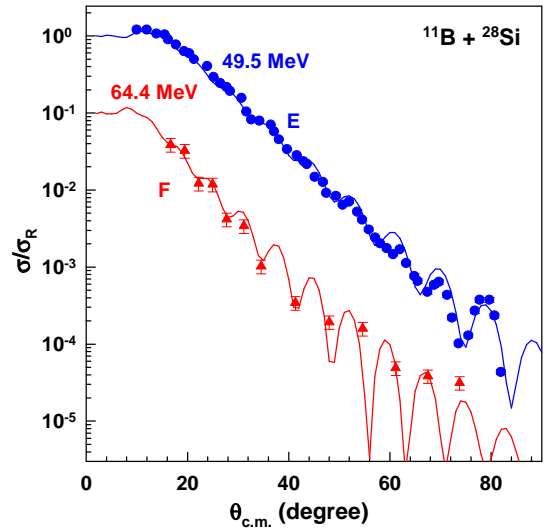}
\end{center}
\caption{(colour online) Elastic scattering angular distributions for $^{11}$B + $^{28}$Si reaction at 49.5 (data taken from Ref.~\cite{park79}) (circle) and 64.4 (triangle) MeV. The solid lines represent the optical model fitting with parameters given in Table~\ref{tbl1}.}
\label{op2}
\end{figure}

\subsubsection{Exit channel}

In addition, the elastic scattering angular distribution for the $^{11}$B + $^{28}$Si reaction has been measured at 64.4 MeV. This system was also studied earlier at different energies \cite{park79} and 49.5 MeV was the highest energy found in the literature. The potential parameters were $V(r)$(36.50, 1.05, 0.74) and $W_F(r)$(29.80, 1.05, 0.81) and $r_C$ = 1.34 fm \cite{park79}. We use these parameters as the starting parameters and following the similar procedure as described earlier the best fit parameters were obtained for the angular distribution at 49.5 MeV (data taken from Ref.~\cite{park79}). These values were taken as the starting parameters to obtain the optical model potential parameters from the elastic scattering angular distribution at 64.4 MeV. The best fit parameters with minimum $\chi^2/N$ values are given in Table~\ref{tbl1} for 49.5 MeV (Set E) and 64.4 MeV (Set F). The strength of the potential and radius values for imaginary component are found to decrease with increasing energy. The fits are shown in Fig.~\ref{op2} by solid lines.

From the above analysis one can find that the optical model potential parameters are very much energy dependent.

\subsection{DWBA calculations}
The theoretical finite range distorted wave Born approximation (DWBA) calculations have been performed using the computer code DWUCK5 \cite{kunz} for all the observed transitions (shown in Fig.~\ref{fig1}) in $^{27}$Al($^{12}$C, $^{11}$B)$^{28}$Si reaction at 73, 81 and 85 MeV. The distorted waves in the entrance and exit channels have been generated using the optical-model potentials. Four sets of optical model parameters have been used in the calculations and they are given in Table~\ref{tbl1}. 

The spectroscopic strength $\bf G$ was extracted for each of the observed transitions using the relation
\begin{equation*}
\left[\frac{d\sigma}{d\Omega} \right]_{exp} = {\bf G} {\rm g} \left[\frac{d\sigma}{d\Omega} \right]_{DW5}
\end{equation*}

\noindent
where $[d\sigma/d\Omega]_{exp}$ is the experimentally measured differential cross-section, 
$[d\sigma/d\Omega]_{DW5}$ is the differential cross-section predicted by the computer code DWUCK5 and $\rm g$ is the light particle spectroscopic strength. For the $^{12}C_{g.s.}$ = $^{11}B_{g.s.} + p$ reaction, the value of the spectroscopic strength is 2.85 \cite{jarc91,jarc92,von}. The spectroscopic strength $\bf G$ is written as
\begin{equation*}
{\bf G} = \frac{2J_f + 1}{2J_i + 1} C^2 S
\end{equation*}

\noindent
where $C^2$ is the isospin Clebsch-Gordan coefficient and $S$ is the spectroscopic factor.

In the present work, the spectroscopic factors are estimated for the ground state and first two low lying states of $^{28}$Si and $^{11}$B nuclei. For all the states the bound state potential parameters are $r$ = 1.29 fm and $a$ = 0.54 fm. The DWBA fits are shown by different curves for different sets in Figs.~\ref{gs}--\ref{ex22}.

\subsubsection{Ground state}
\begin{figure} [h] \begin{center}
\epsfysize=9.5cm
\epsffile[47 51 471 761]{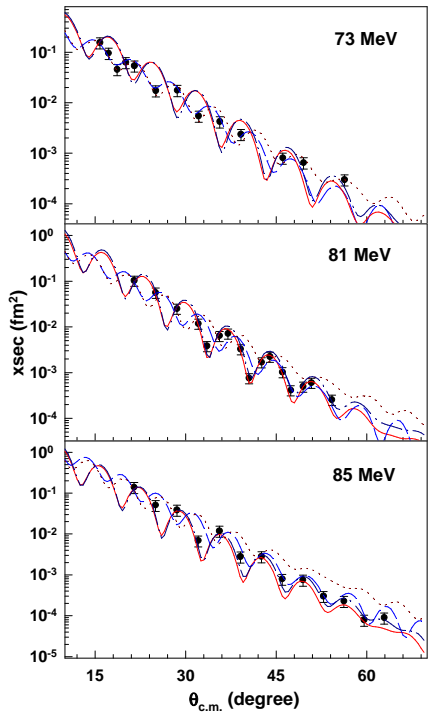}
\end{center}
\caption{(colour online) Angular distributions of differential cross-section for the 0.0 MeV 0$^+$ transition in $^{27}$Al($^{12}$C, $^{11}$B)$^{28}$Si reaction at 73, 81 and 85 MeV. The dashed, dotted, solid and dash-dotted curves are the results of DWBA calculations using optical model parameter Set A-E, B-F, C-F and D-F, respectively. }
\label{gs}
\end{figure}

The angular distributions of the transition corresponding to the ground state (0$^+$; 0.0 MeV) peak have been displayed in Fig.~\ref{gs} for all three bombarding energies. The spin and parity of the ground state of $^{28}$Si nucleus are well established \cite{endt}. The $l$-transfer is equal to 3 for this level. The ground state configuration of the exit channel is $\pi$($1p_{3/2}{}^{-1},1d_{5/2}$) in the shell model, $i.e.$, one proton from $1p_{3/2}$ level in $^{12}$C nucleus transferred to $1d_{5/2}$ level in $^{28}$Si nucleus. Both the nuclei are in ground state. The finite range DWBA calculations have been done with a $\Delta l$ = 3 transfer using the optical model parameters obtained from the elastic scattering analysis (Table~\ref{tbl1}). Eighty-five partial waves were used and the integration was performed in step intervals of 0.05 fm from 0.0 to 30 fm. The results are the same when ninety-five partial waves were used and the integration was performed in steps of 0.1 fm from 0.0 to 40 fm. Coulomb excitation was found to be important at all the incident energies. 

It has been found from Fig.~\ref{gs} that calculation with set A-E fits the experimental angular distribution at 73 MeV (dashed curve) except at higher angels ($\theta_{c.m.}$ $\geq$ 45$^o$). The calculation with parameter set B-F fits the data very well (dotted curve). However, the calculations with sets C-F (solid curve) and D-F (dash-dotted curve) are about 3$^o$ out of phase with the data at 73 MeV.

The experimental angular distribution at 81 MeV has been fitted well with the DWBA calculation using set C-F (solid curve). The calculation using set D-F fits the data except at higher angles ($\theta_{c.m.}$ $\geq$ 40$^o$). On the other hand, the DWBA calculation using set D-F fits (dash-dotted curve) the experimental angular distribution at 85 MeV very well. Except at higher angles ($\theta_{c.m.}$ $\geq$ 45$^o$), the calculation with set C-F also fits the data. For both 81 and 85 MeV bombarding energy, the calculations with set A-E and B-F are about 3$^o$ out of phase with the data.

The spectroscopic strength and the spectroscopic factor deduced from strength are given in Table~\ref{tbl2} for all sets of potential parameters. 

\begin{table} [h]
\caption{Spectroscopic factors deduced for ground state 0$^+$, 0.0 MeV. The transferred nucleon is in 1d$_{5/2}$ level. The shell model predicted strength is 0.53 \cite{das99,wil73}.}
\begin{tabular}{lccccccc} \hline\hline
&\multicolumn{3}{c}{$\bf G$ }&&\multicolumn{3}{c}{$S$}  \\
Set&73 &81 &85&&73 &81 &85  \\ 
&(MeV)&(MeV)&(MeV)&&(MeV)&(MeV)&(MeV)  \\ \hline \hline
A-E&1.8&4.0&7.0&&2.40&5.33&9.33 \\
B-F&1.2&2.5&4.0&&1.60&3.33&5.33 \\
C-F&0.5&1.0&1.06&&0.67&1.41&1.33 \\
D-F&0.46&1.0&1.0&&0.80&1.33&1.33 \\   \hline \hline
\end{tabular}
\label{tbl2}
\end{table}

\subsubsection{Excited states of $^{28}$Si}
\begin{figure} [h] \begin{center}
\epsfysize=9.5cm
\epsffile[47 51 472 764]{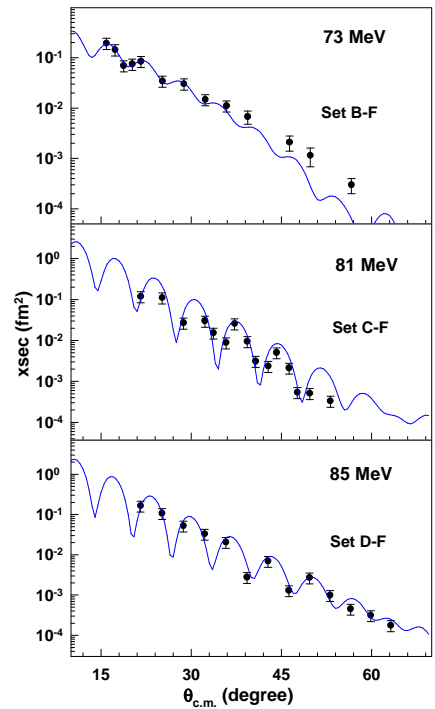}
\end{center}
\caption{(colour online) Angular distributions of differential cross-section for the 1.78 MeV 2$^+$ transition in $^{27}$Al($^{12}$C, $^{11}$B)$^{28}$Si reaction at 73, 81 and 85 MeV. The solid curves represent the DWBA fit with optical model potential parameters (see text).}
\label{ex1}
\end{figure}

\begin{figure} [h] \begin{center}
\epsfysize=9.5cm
\epsffile[47 51 471 761]{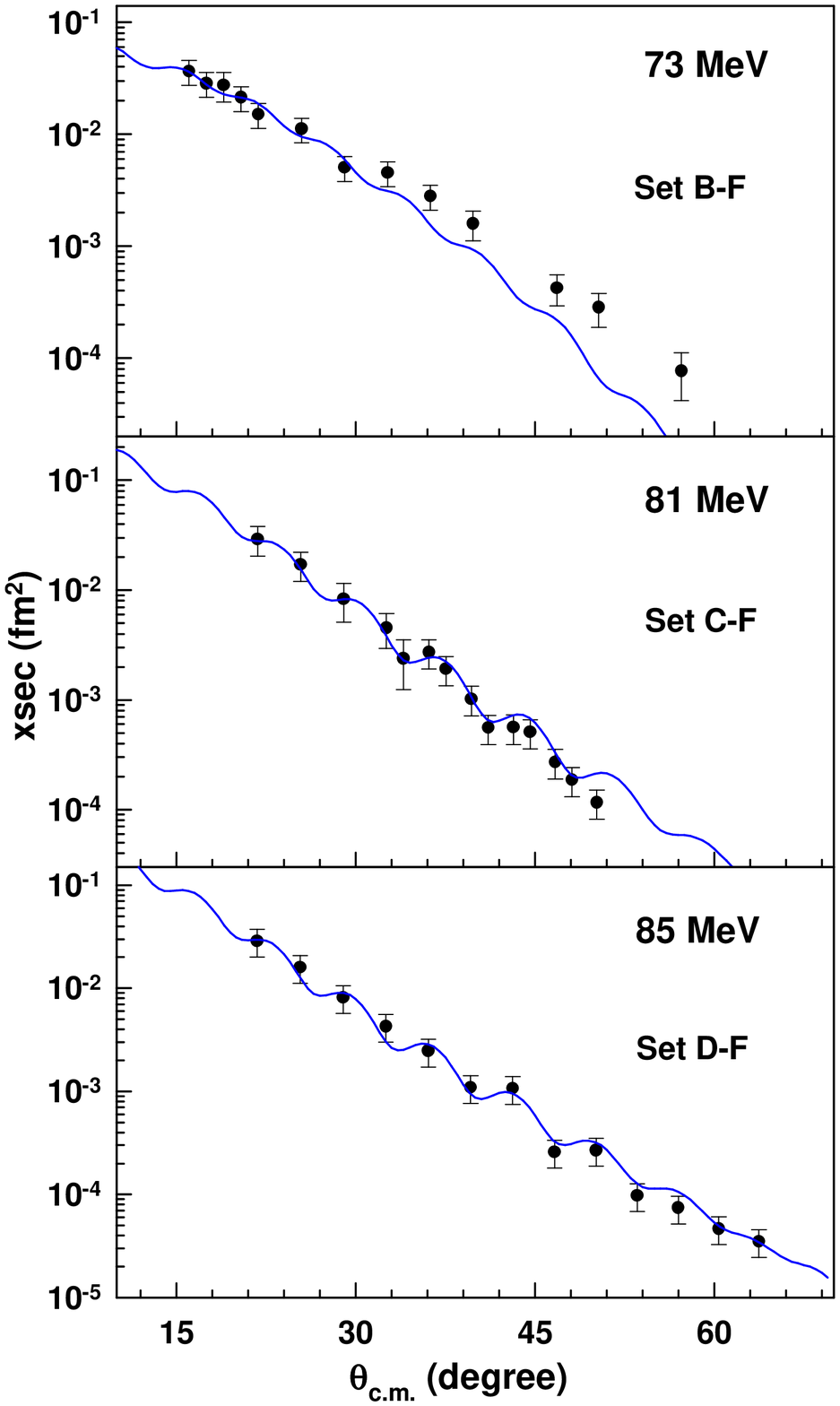}
\end{center}
\caption{(colour online) Same as Fig.~\ref{ex1} for the 4.62 MeV 4$^+$ transition. }
\label{ex2}
\end{figure}

The angular distributions of two transitions corresponding to the first excited state (2$^+$; 1.78 MeV) and second excited state (4$^+$; 4.62 MeV) of $^{28}$Si have been displayed in Figs.~\ref{ex1}~and~\ref{ex2}. The spin and parity of the $^{28}$Si nucleus are well established \cite{endt}.

The angular distributions corresponding to the first excited state peak are displayed in Fig.~\ref{ex1} for all the incident energies. The $l$-transfer is equal to 1 for this level. The excited state may have two configuration; $\pi$($1p_{3/2}{}^{-1},2s_{1/2}$) or $\pi$($1p_{3/2}{}^{-1},1d_{3/2}$) in the shell model. The transferred proton from $1p_{3/2}$ level may go to $2s_{1/2}$ level or $1d_{3/2}$ level and produce the first excited state of $^{28}$Si nucleus. Therefore the experimental angular distribution has contribution from both the configurations. It has been found that the ratio $2s_{1/2}$:$1d_{3/2}$ is equal to 0.9:0.1 at 73 MeV and goes to 0.85:0.15 at 85 MeV. The finite range DWBA calculations with optical model parameter set B-F, C-F and D-F fit the experimental data at 73, 81 and 85 MeV, respectively. The DWBA fits are shown by solid curves in Fig.~\ref{ex1}. The deduced spectroscopic strengths and factors are given in Table~\ref{tbl3}.

The experimental angular distributions corresponding to the second excited state peak are shown in Fig.~\ref{ex2}. Here, the $l$-transfer is $l = 1$ and the exit channel configuration is $\pi$($1p_{3/2}{}^{-1},1d_{3/2}$) in the shell model. The excited proton goes to the $1d_{3/2}$ level and produce the 4$^+$ state of $^{28}$Si nucleus. The DWBA calculations have been done and the fits are shown in Fig.~\ref{ex2} by solid curves. The spectroscopic strengths deduced from the angular distributions are given in Table~\ref{tbl3} for all three bombarding energies. The data at 73, 81 and 85 MeV have been fitted with finite range DWBA calculations using the optical model parameter set B-F, C-F and D-F, respectively.

\subsubsection{Excited states of $^{11}$B}

\begin{figure} [h] \begin{center}
\epsfysize=9.5cm
\epsffile[47 50 470 759]{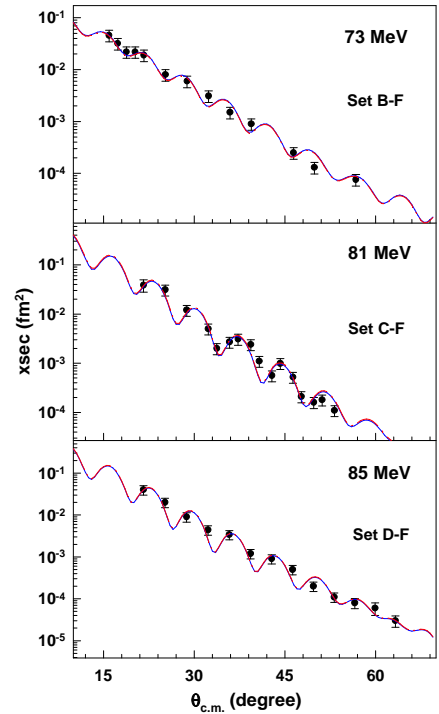}
\end{center}
\caption{(colour online) The angular distributions for the 2.12 MeV 1/2$^-$ transition in $^{27}$Al($^{12}$C, $^{11}$B)$^{28}$Si reaction at 73, 81 and 85 MeV. The DWBA fits are shown by solid and dash-dotted curves (see text).}
\label{ex12}
\end{figure}

\begin{figure} [h] \begin{center}
\epsfysize=9.5cm
\epsffile[47 50 470 759]{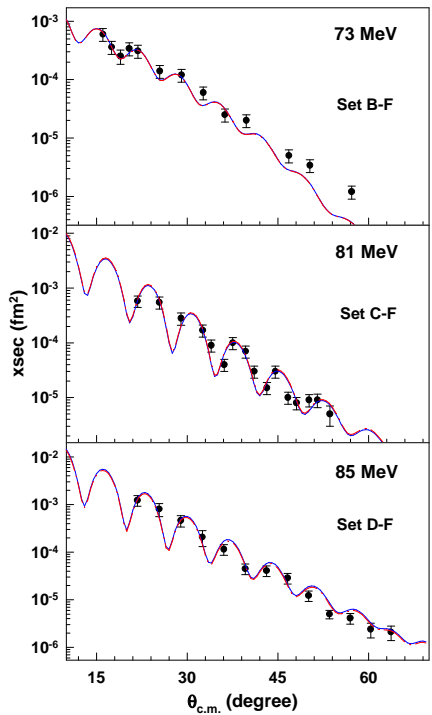}
\end{center}
\caption{(colour online) Same as Fig.~\ref{ex12} for the 4.44 MeV 5/2$^-$ transition.}
\label{ex22}
\end{figure}

The angular distributions of two transitions corresponding to the first excited state (1/2$^-$; 2.12 MeV) and second excited state (5/2$^-$; 4.44 MeV) of $^{11}$B have been displayed in Figs.~\ref{ex12}~and~\ref{ex22}. The spin and parity of the $^{11}$B nucleus are well established \cite{endt}. In this case, the excitation energy of the total system is such that the transition leading to residual nucleus ($^{28}$Si) in its ground state should favour the excitation of the first few excited states in ejectile ($^{11}$B) due to the simultaneous fulfilment of the momentum matching conditions deduced from the semiclassical analysis of the reaction \cite{po75}. 

The experimental angular distributions corresponding to the first excited state peak are displayed in Fig.~\ref{ex12} for all the incident energies. The $l$-transfer is equal to 3 for this level. Here the proton may transfer from $^{12}$C ground state to $^{28}$Si ground state ($1d_{5/2}$ level) and leaving the $^{11}$B nucleus in its first excited state in $1p_{1/2}$ level. Therefore the exit channel configuration in shell model is $\pi$($1p_{1/2}^{-1}1d_{5/2}$). The spectroscopic strength for the $^{12}C_{g.s.}$ = $^{11}B_{2.12} + p$ reaction is 0.75 \cite{jarc91} and/or 0.64 \cite{jarc92}. The deduced spectroscopic strengths and factors are given in Table~\ref{tbl3}. 

\begin{table*} 
\caption{Spectroscopic factors deduced for excited states of $^{28}$Si and $^{11}$B.}
\begin{tabular}{cccccccc} \hline\hline
$E_X$&$J^{\pi}$&Level&Energy&Set&$\bf G$&G&$S$ \\
(MeV)&&&(MeV)&&(Present work)&Shell model&(Present work) \\ \hline \hline
\multicolumn{8}{c}{$\bf ^{28}Si$} \\
1.78&2$^+$&2s$_{1/2}$+1d$_{3/2}$&73&B-F&0.86, 0.10&0.38, 0.06 \cite{das99,wil73}&3.44, 0.20  \\ 
&&&81&C-F&0.90, 0.10&&3.60, 0.20  \\
&&&85&D-F&0.67, 0.08&&2.68, 0.16  \\ 
&&&&&&& \\
4.62&4$^+$&1d$_{3/2}$&73&B-F&2.20&0.33 \cite{das99,wil73}&4.40 \\ 
&&&81&C-F&0.95&&1.90  \\
&&&85&D-F&0.90&&1.80  \\  \hline 
\multicolumn{8}{c}{$\bf ^{11}B$} \\
2.12&1/2$^-$&1p$_{1/2}$&73&B-F&1.0, 1.2&0.75 \cite{jarc91}, 0.64 \cite{jarc92}&0.67, 0.80 \\ 
&&&81&C-F&0.85, 0.95&&0.57, 0.63  \\
&&&85&D-F&0.70, 0.80&&0.47, 0.53 \\  
&&&&&&& \\
4.44&5/2$^-$&1d$_{5/2}$&73&B-F&0.71, 0.50&0.55 \cite{jarc91}, 0.79 \cite{jarc92}&1.42, 1.0 \\ 
&&&81&C-F&0.39, 0.26&&0.78, 0.52  \\
&&&85&D-F&0.42, 0.31&&0.84, 0.62 \\  \hline \hline
\end{tabular}
\label{tbl3}
\end{table*}

The angular distributions corresponding to the second excited state peak are shown in Fig.~\ref{ex22}. Here, the $l$-transfer is equal to 3 and the exit channel configuration is $\pi$($1d_{5/2}^{-1}1d_{5/2}$) in the shell model. The transferred proton goes to the $1d_{5/2}$ level $i.e.$, ground state of residual nucleus $^{28}$Si leaving the second excited state of $^{11}$B in $1d_{5/2}$ level. The light particle spectroscopic strength for the $^{12}C_{4.44}$ = $^{11}B_{g.s.} + p$ reaction is 0.55 \cite{jarc91} and/or 0.79 \cite{jarc92}. The spectroscopic strengths and factors deduced from the angular distribution are given in Table~\ref{tbl3} for all the incident energies. 

The finite range DWBA calculations have been done using the optical model parameter set B-F, C-F and D-F for the angular distributions at 73, 81 and 85 MeV, respectively, for both the excited level of $^{11}$B nucleus. In this case, two different values have been found in literature \cite{jarc91,jarc92} for the light particle psectroscopic strength (g). In this analysis, we consider both the values and extract two different values for scpectroscopic strength as well as spectroscopic factor. The DWBA fits are shown in Figs.~\ref{ex12}~and~\ref{ex22} by solid (g value taken from Ref.~\cite{jarc91}) and dash-dotted (g value taken from Ref.~\cite{jarc92}) curves. The fits are overlapping, however, we found two values for spectroscopic factor.

\section{Summary and Conclusion}

The one proton transfer reaction channel has been studied using the $^{27}$Al($^{12}$C, $^{11}$B)$^{28}$Si reaction at three bombarding energies at 73, 81 and 85 MeV. In $^{11}$B fragment spectrum, different peaks have been identified as the transitions from ground state and excited states of $^{11}$B and $^{28}$Si nuclei. The experimental angular distributions of ground (0$^+$; 0.0 MeV), first (2$^+$; 1.78 MeV) and second (4$^+$; 4.62 MeV) excited states of $^{28}$Si as well as first (1/2$^-$; 2.12 MeV) and second (5/2$^-$; 4.44 MeV) excited states of $^{11}$B have been analysed using the finite range DWBA computer code DWUCK5. The bound state potential for proton has Wood-Saxon form and the parameters are $r$ = 1.29 fm and $a$ = 0.54 fm for all the transitions. The optical model potential has the parametric Wood-Saxon form for both real and imaginary potentials. The optical model potential have been derived for $^{12}$C + $^{27}$Al reaction at four energies and for $^{11}$B + $^{28}$Si reaction at two energies in the present work and the potential parameters are found to be energy dependent. 

The potential parameter Sets A, B, C and D for entrance channel and Sets E and F for exit channel have been derived at different incident energies. The finite range DWBA calculation using potential parameter set A-E and B-F are in phase, however, the amplitudes are different at higher angles ($\theta_{c.m.}$ $\geq$ 40$^o$). On the other hand, the finite range DWBA calculation using potential parameter set C-F and D-F are in phase, with a slight difference in amplitudes at higher angles ($\theta_{c.m.}$ $\geq$ 40$^o$). There is 3$^o$ phase difference between the calculations using sets A-E, B-F and C-F, D-F. Experimentally, this phase difference is also observed between the angular distributions at 73 MeV and 81 MeV. From Fig.~\ref{gs} and Table~\ref{tbl2}, we can say that the optical model parameters should be derived at same and/or nearby energy to avoid the effect of energy dependence of optical model potential on the extracted spectroscopic factors. 

\acknowledgments
The authors thank the Pelletron operating staff for smooth running of the machine. They also thank the personnel at TIFR for supplying the P10 Gas cylinder during the experiment.

\end{document}